\documentclass[final]{iopart}
\usepackage{setspace}
\usepackage{iopams}  
\usepackage{verbatim}
\usepackage{graphicx}
\usepackage{rotating}
\usepackage{mathrsfs}
\newcommand{\scri}{{\mathscr I}}

\begin{document}

\title{Analytic conformal compactification of Schwarzschild spacetime}

\author{Jakub Hal\'a\v{c}ek and Tom\'a\v{s} Ledvinka}

\address{Institute of Theoretical Physics, Faculty of Mathematics and Physics, Charles University in Prague, V Hole\v{s}ovi\v{c}k\'ach 2, 180 00 Praha 8, Czech Republic}
\ead{j.halacek@gmail.com {\rm and} tomas.ledvinka@mff.cuni.cz}
\begin{abstract}
Among the coordinates used to construct a conformal compactification of the Schwarzschild spacetime, 
none of them simultaneously extend smoothly both through an event horizon and beyond null infinity.
To construct such coordinates, instead of starting with the Kruskal-Szekeres coordinates we assume
direct analytic transformation between Schwarzschild and compactified coordinates and determine their behavior on the event horizon and at null infinity.
We then propose an example of such coordinates and illustrate the way they cover the conformally extended Schwarzschild spacetime as well as their suitability for numerical applications.
\end{abstract}

\pacs{04.20.-q, 97.60.Lf, 04.25.D-j}

\section{Introduction}
A very useful tool used to discuss various aspects of black-hole spacetimes are the so-called Carter-Penrose diagrams, which 
are based on the mathematical concept of conformal maps between Lorentzian manifolds. Such maps preserve causal structure, but they fit 
the whole spacetime into a finite picture and so they can be useful to illustrate the physical properties of spacetimes such as the structure of horizons or the global shape of worldlines of various observers and null particles, etc. 

The standard depiction of compactified Schwarzschild spacetime given in textbooks \cite{MTW,wald,novikov} simply shrinks the well-known Kruskal-Szekeres
construction of maximal extension of the Schwarzschild metric into a finite picture. If this depiction appears next 
to a diagram of compactified Minkowski spacetime (see Figure \ref{fig_ck}), the two plots seem not only to differ at regions close to the horizon, 
but also near the null and spatial infinities. This is not supposed to happen as the Schwarzschild spacetime 
is the most famous member of the family of spacetimes with Minkowski-like infinities -- the so-called asymptotically flat spacetimes. 

Since the compactified diagrams of the Schwarzschild black-hole spacetime are regularly used to describe processes from the point of view of very distant observers 
or, e.g., to illustrate geometrical objects stretching to infinity, an important distortion may appear.
Typical example can be found in \cite{zenginoglu}, where hyperboloidal slices both in Minkowski and in Schwarzschild spacetimes are studied (see \cite{zenginoglu}, Figure 4 and 10).
The fact, that curves representing the slices look so differently near the null infinity is only an artifact due to very different behavior of each compactification near infinity.
Similarly, also the depiction of the spacelike slicing of the Schwarzschild geometry represented by hypersurfaces $t=\rm const.$ is distorted at infinity (see Figure \ref{fig_ck}). 
Thus the standard compactification of the Schwarzschild spacetime cannot, for example, be used to illustrate faithfully the geometry of slices through the spacetime with moving Schwarzschild black hole, because the angle under which slices meet at spatial infinity no longer indicates their relative velocity.
Practical problems may also arise when such coordinates are used numerically, since they do not behave well near infinities.

Since both the Schwarzschild geometry and the Carter-Penrose diagrams represent usual textbook topics, several coordinate transformations which should amend this distortion at infinities have been proposed \cite{novikov,hervik}. Even though the existence of analytic coordinates on the manifold conformally-related to the Schwarzschild spacetime that cover both horizon and null infinities has been proven \cite{friedrich2003}, the available closed-form transformations given in \cite{novikov,hervik} are not analytic in the null infinity. 
This situation is not limited to the Schwarzschild spacetime -- in \cite{podolsky} 
the dissimilarity of asymptotic regions between available Carter-Penrose diagrams for asymptotically flat exact solutions of Einstein equations and those for the 
Minkowski spacetime is mentioned be a general feature. 
 
In this article we propose coordinates which provide an analytical map on the compactified manifold and thus lead to the Carter-Penrose diagram of the Schwarzschild spacetime 
much more similar to the Carter-Penrose  diagram of compactified Minkowski spacetime. 
In the following Section \ref{SecStd} we first review the available compactification transformations. Then, in Section \ref{SecNew}, we show how to make the compactification transformation analytical at null infinity and horizon.
In the final section we discuss the obtained Carter-Penrose diagrams and the implications of the analytic properties of transformations in numerical applications.
Well-known properties of the compactification transformation for the Minkowski spacetime, and for asymptotically flat spacetimes in general, are summarized in the \ref{AppA}.

\section{Standard compactification of Kruskal-Szekeres coordinates}
\label{SecStd}
The usual compactified diagram of the Schwarzschild spacetime is constructed from the null Kruskal's coordinates using the transformation similar to (\ref{minkkompakt}) of Minkowski spacetime.
Let us recall, that the radial null geodesic for the Schwarzschild metric 
\begin{equation}
 ds^2 = -\left(1-\frac{2M}{r}\right) dt^2 +\frac{dr^2}{1-\frac{2M}{r}}  + r^2 d\omega^2
\end{equation}
from which the Kruskal's coordinates are derived are given by the partially implicit prescription
\begin{eqnarray}
    f(r)  &= V-U,~
    t &=  U+V,~
    \theta = {\rm const.},~\phi = {\rm const.}
\end{eqnarray}
where $r_* = f(r)$ is the so-called tortoise coordinate,
\begin{equation}
f(r) = r + 2M \ln \left( \frac{r}{2M} - 1 \right).
\label{torto}
\end{equation}
The outgoing radial null geodesics are parametrized by $V$  and labeled by $U, \theta, \phi=\rm const.$, the ingoing radial null geodesic are parametrized by $U$.

\begin{figure}[h]
\begin{center}
\includegraphics[width=\linewidth]{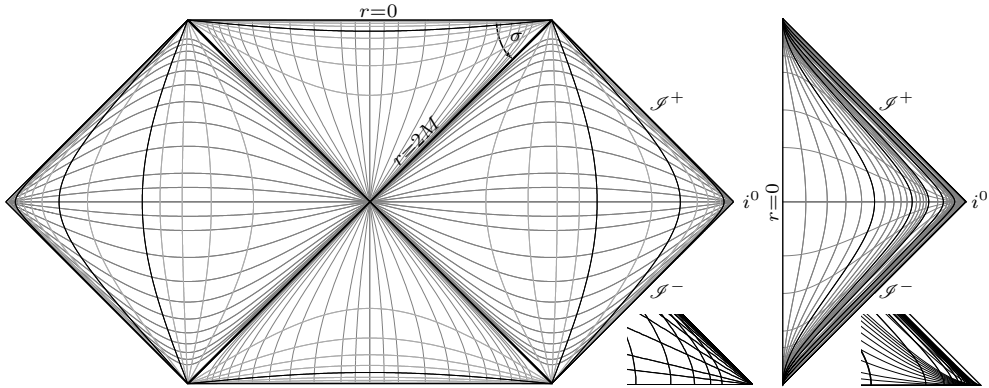} 
\end{center}
\caption{The standard compactified Kruskal diagram of Schwarzschild spacetime (left) and Carter-Penrose diagram of Minkowski spacetime (right). 
For $M=1$ coordinate grid of both corresponds to $r=0,0.2,0.4,...,3,4,5,6,...,10,20,...$ and $|t|=0,0.5,1,2,...$. Emphasized are lines $r=0,1,2,3,5,10$. 
The details of regions near $i^0$ are also displayed.}
\label{fig_ck}
\end{figure}

To achieve the conformal compactification we need to construct some mapping which will
put the infinities $U\rightarrow \infty \vee V\rightarrow -\infty$ (horizon)
and $U\rightarrow -\infty \vee V\rightarrow +\infty$ (the null infinity) into the inner points of some larger, unphysical manifold. While later we will show that to obtain an analytical compactification of the Schwarzschild spacetime it is easier to treat at once both the null infinity and the horizon, the standard compactification
uses the function $\ln(x)$ to penetrate the horizon first \cite{kruskal} and only then 
the function $\tan(x)$ is plugged in to fit the Kruskal coordinates into a finite interval \cite{MTW}.
These new coordinates ${\cal U},{\cal V}\in (-\pi/2,\pi/2)$
are related to Schwarzschild coordinates $t, r$ by the transformation
\begin{eqnarray}
\label{g_MTW}
    \label{rcompkru}
    f(r({\cal U},{\cal V}))  &= 2M \ln \left( \tan\;{\cal V}\right)+2M \ln \left( -\tan\; {\cal U} \right)~,\\
    t({\cal U},{\cal V}) &=  \Re \left[ 2M \ln \left({ \tan\;{\cal V}} \right)-2M \ln \left( -\tan\;{\cal U} \right)\right].\label{kruskaltime}
\end{eqnarray}
To point out the structure of this transformation, we prefer to use the real part on the right-hand side of (\ref{kruskaltime}) instead of the
usual $\ln|\tan\; {\cal U}/\tan\; {\cal V}|$.
Then
\begin{equation}
 ds^2 = -\frac{1}{\cos^2 {\cal U} \cos^2 {\cal V}}\frac{32 M^3}{r} e^{-\frac{r}{2M}} d\,{\cal U}\, d{\cal V}+r^2 d\omega^2~.
\label{dsuv_1}
\end{equation}
When $r<2M$ the transformation (\ref{rcompkru}) becomes a complex-valued implicit function prescription 
for a real function of two real variables $r({\cal U},{\cal V})$. 
Because the metric coefficients in (\ref{dsuv_1}) are simple analytic functions of ${\cal U}, {\cal V}$ and $r$,
the fact that the (here compactified) Kruskal maximal extension can be found is implied
by the existence of analytic function $r({\cal U},{\cal V})$ solving (\ref{rcompkru}).
Since the Lambert function $W_0(z)$ appearing in the transformation
\begin{equation}
r({\cal U},{\cal V}) = 2M W_0(-e^{-1} \tan\;{\cal U} \tan\;{\cal V} )
\end{equation}
is analytic on the real axis along the interval $z\in (-e^{-1},\infty)$ \cite{lambert},
the line element (\ref{dsuv_1}) is analytic in the domain where $-\infty < \tan\;{\cal U} \tan\;{\cal U} < 1$ i.e. $r>0$, namely at the horizon. 
The way the Schwarzschild coordinates $r, t$ cover this domain can be seen in Figure \ref{fig_ck}.

It is known that the exponential term $e^{-r/2M}$ in the Kruskal line element (\ref{dsuv_1}) prohibits one to satisfy conditions (\ref{dOmega}-\ref{ddOmega}) required for asymptotically flat spacetimes. 
Thus  several coordinate transformations have been given in the literature \cite{novikov,hervik} which should lead to compactification behaving the same way as the Minkowski spacetime for $r\rightarrow\infty$. Namely, as the adjective conformal should mean angle-preserving, 
these coordinates restore the way the curves representing hypersurfaces  $t=\rm const.$ behave near $i^0$. 
As an example let us check the properties of the transformation proposed in \cite{novikov}
\begin{equation}
{\mathscr U} = \arctan {\rm argsinh}\;{\cal U},~~{\mathscr V} = \arctan {\rm argsinh}\;{\cal V},
\label{UVNFdef}
\end{equation}
which leads to an implicit equation for $r({\mathscr U},{\mathscr V})$ given by
\begin{equation}
f( r({\mathscr U},{\mathscr V}) ) = 2M \ln \left(  2\sinh \tan\;{\mathscr V}\right)+2M \ln \left( -2  \sinh \tan\;{\mathscr U} \right) .
\label{fruv2}
\end{equation}
Using $\Omega_2 = {\cos{\mathscr U}\; \cos {\mathscr V}}/({16 M^2})$
the conformally related metric reads 
\begin{equation}
 \widetilde{ds}{}^2 = - \frac{\coth \tan  {\mathscr U}\; \coth \tan {\mathscr V}}{16 M^2}\left(1-\frac{2M}{r_2}\right)  {d{\mathscr U}}\; {d{\mathscr V}} + \Omega_2^2 r^2 d\omega^2.
\label{dsuv_2k}
\end{equation}
With this metric the conformal factor $\Omega_2$ satisfies conditions (\ref{dOmega}-\ref{ddOmega}), but the metric still cannot be extended through $\scri^\pm$ since the function $\coth \tan  z$ has an essential singularity at $z=\pm \pi/2$ which simplifies to a discontinuity for real arguments. Similarly the transformation mentioned in \cite{hervik}
which is also based on Kruskal coordinates provides compactification bounded by $\scri$. 
The fact that the conformally related metric is not available beyond $\scri^\pm$ may seem to be only a small issue but later we will show that this fact 
can be numerically observed  from the behavior of metric coefficients within the physical domain $2M<r<\infty$ (see Figure \ref{fig_hypsurf}). 

\section{Analytic conformal compactification of Schwarzschild spacetime}
\label{SecNew}
We do not search for the `right' compactification of Kruskal coordinates, instead we propose the direct transformation between Schwarzschild coordinates $t,r$ and the compactified coordinates $u,v$ in the form
\begin{eqnarray}
f( r(u,v) ) &= h(v)+h(-u),
\label{fruv3}
\\
t(u,v) &= h(v) - h(-u).
\end{eqnarray}
This choice is inspired by the common form of transformations (\ref{rcompkru}) and  (\ref{fruv2}), which is given by the fact that such a transformation changes $-dt^2+dr_*^2$ into a term proportional to $du\,dv$.
The choice of parameters of a yet unknown function $h$ is such that in the exterior Schwarzschild region $r>2M$ both $v$ and $-u$ are positive.
Then the Schwarzschild line element reads
\begin{equation}
 \overline{ds}^2 = - 4 \left(1-\frac{2M}{r}\right)h'\!(v)\,h'\!(-u)~{du}\, {dv} + r^2~d\omega^2.
\label{dsuv_3}
\end{equation}
Let us now discuss what kind of function $h$ would lead to an analytic conformal embedding (\ref{dsOmegads}) of the complete Schwarzschild manifold using the conformal factor 
$\Omega_3 \sim \cos u \cos v$, which implies we assign $u = -\pi/2, v=\pi/2$ to coordinates of null infinities $\scri^\pm$ and the horizon $\cal H^\pm$ is simply put at $u=0~ \vee~ v=0$.
The transformation (\ref{fruv3}) again represents a complex-valued implicit equation for the real function $r(u,v)$. 
To prescribe precisely the behavior of the transformation function $h(x)$ in the complex domain, we decompose it into 
\begin{equation}
 h(x) = \alpha(x)+2M\ln \beta(x),
\label{gansatz}
\end{equation}
where $\alpha, \beta$ are analytic functions on $(-\pi,\pi)$ up to simple poles at $\pm \pi/2$ (i.e. at $\scri^\pm$, see eq.(\ref{impeqpsi})). 
Indeed, $\alpha$ and $\beta$ are also restricted by the fact that for a regular transformation $h'(x)\ne 0$ on this interval.

From (\ref{torto}) we see that $\beta$ should become negative for $r<2M$ so that imaginary parts on both sides of (\ref{fruv3}) match
due to the common factor $2M$ in front of the logarithm in (\ref{gansatz}) and (\ref{torto}). The solution of (\ref{fruv3}) can be given using the Lambert function
\begin{equation}
 r = 2M \left[1+W_0\left( \beta(v) \beta(-u) e^{\frac{\alpha(v)+\alpha(-u)}{2M}-1} \right)  \right],
 \label{rW0}
\end{equation}
which (using $W_0(x)\sim x$) implies that analytic covering of the horizon requires 
\begin{equation}
\alpha(x) \sim 1,~\beta(x)\sim x~~{\rm for}~~x\simeq 0~. 
\label{hbeh0}
\end{equation}

We have to further restrict $\alpha, \beta$ so that conformally related metric $\widetilde g$ (primarily its metric component $\widetilde g_{\theta\theta}$) is analytic near ${\scri}^\pm$.
Eq. (\ref{rW0}) cannot be used directly as we would need to regularize expression $0.W(\infty e^\infty)$.
We rather decompose $\sqrt{\widetilde g_{\theta\theta}}=r \Omega$ into a sum of two terms 
\begin{equation}
 r\Omega = [ \alpha(v) + \alpha(-u) ]\,\Omega ~+~  [r-\alpha(v)- \alpha(-u)]\,\Omega 
\label{rOmegaDekompo}
\end{equation}
and then we require both of them to be analytic functions. This for the first term simply yields
$\alpha(v)\sim 1/\Omega$ near ${\scri}^+$ and  $\alpha(-u)\sim 1/\Omega$ near ${\scri}^-$. 
The analytic properties of the second term on the right-hand side of (\ref{rOmegaDekompo}) are for an analytic conformal factor $\Omega$ equivalent to the properties of an auxiliary function $\psi(u,v)$ we define by the relation 
\begin{equation}
 2M \left[ 1+\psi(u,v)\right] \equiv r-\alpha(v)-\alpha(-u) = 2M\ln \frac{\beta(v)\beta(-u)}{\frac{r}{2M}-1}~.
\label{defofpsi}
\end{equation}
We can rewrite this definition as an implicit equation
\begin{equation}
 \psi(u,v) = -1-\ln \frac{ \frac{1}{\alpha(-u)}+\frac{1}{\alpha(v)}+2M\frac{1}{\alpha(-u)}\frac{1}{\alpha(v)}\psi(u,v) }{2M \frac{\beta(-u)}{\alpha(-u)}\frac{\beta(v)}{\alpha(v)}} ~.
\label{impeqpsi}
\end{equation}
If we consider e.g. $\scri^+$ where $1/\alpha(v) \rightarrow 0$ as $v \rightarrow \pi/2$, the right-hand side of (\ref{impeqpsi}) does not depend on $\psi$ there and  (\ref{impeqpsi}) thus explicitly determines values of $\psi(u,v=\pi/2)$. Indeed, we also need that the ratio $\beta(v)/\alpha(v)\sim 1$ for $v \rightarrow \pi/2$, i.e. the poles of both functions must cancel out:
\begin{equation}
\alpha(x) \sim \beta(x)\sim \frac{1}{\cos x}~~{\rm for}~~x\simeq \frac{\pi}{2}~. 
\label{hbehpi2}
\end{equation}
With this behavior of $\alpha$ and $\beta$ at $\scri^\pm$ both sides of (\ref{impeqpsi}) are guaranteed to have different derivatives with respect to $\psi$ and the implicit function theorem then implies that $\psi$ is analytic function at ${\scri}^\pm$. No further restrictions on $\alpha$ and $\beta$ are implied by a regularity of the metric component $\widetilde g_{uv}$.

\begin{figure}[t]
\begin{center}
\includegraphics[width=\linewidth]{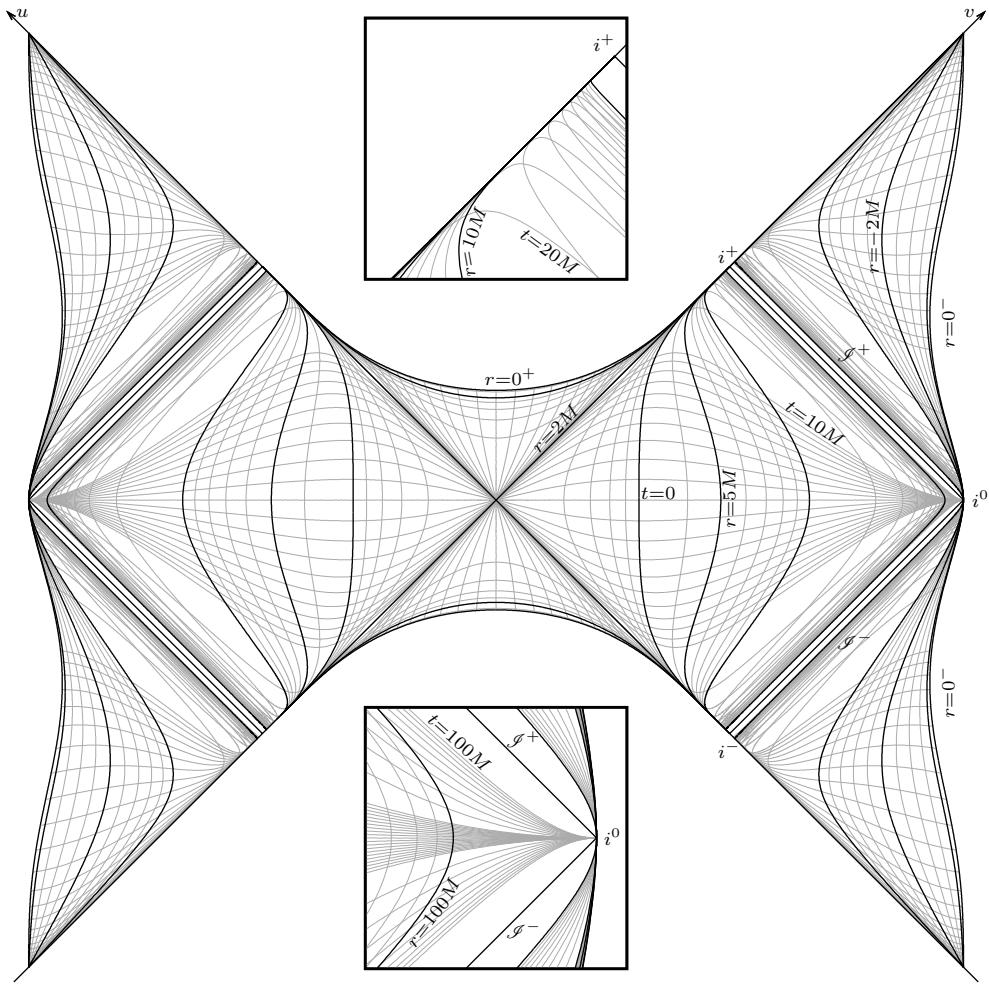} 
\end{center}
\caption{Carter-Penrose diagram of Schwarzschild spacetime and its analytic extension beyond $\scri^\pm$ using coordinates (\ref{our_gx}). 
Compactification scale $\mu=3M$ is used. 
Coordinate grid corresponds to $r/M=0,0.5,1,1.2,...,2.8,3,4,5,6,...,10,20,...,100,\infty,-100,-90,...,-10,-9,...,-1,0$ and $|t/M|=0,1,...,10,20,...,100$. 
Emphasized are lines $|r/M|=0,1,2,3,5,10,100$. 
  }
\label{fig_cps}
\end{figure}

\begin{figure}[h]
\begin{center}
\includegraphics{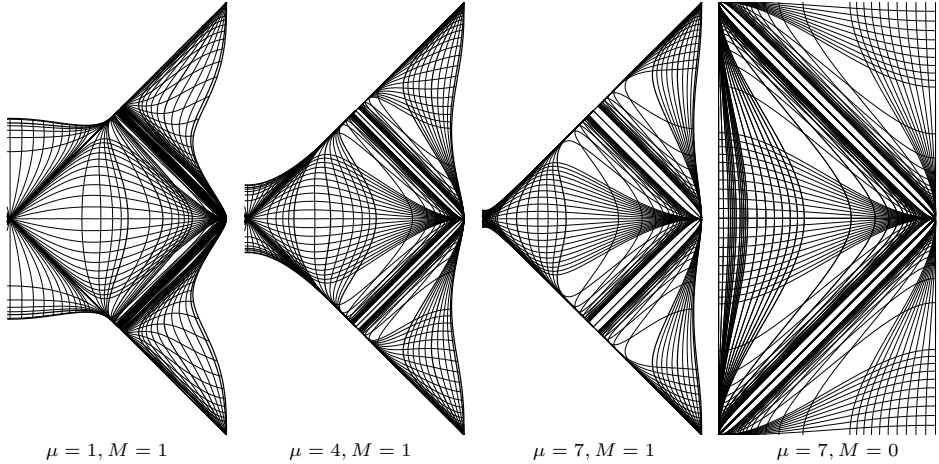}\\
\scriptsize$\mu=1, M=1$~~~~~~~~~~~~~~~~~$\mu=4, M=1$~~~~~~~~~~~~~~~~~$\mu=7, M=1$~~~~~~~~~~~~~~~~~$\mu=7, M=0$
\end{center}
\caption{Carter-Penrose diagram of Schwarzschild spacetime with three different choices of the compactification scale $\mu$.
A compactified Minkowski spacetime is shown on the right panel. 
To construct this diagram the transformation (\ref{minkkompakt}) is plotted with the factor $2$ there replaced by $1/\mu$ so that $\mu=7$ can be used to make the outer regions of the two right panels match.}
\label{fig_compacscale}
\end{figure}

We found a reasonably simple choice of function
\begin{equation}
 h(x) = \frac{\mu}{\cos x} + 2M\ln \frac{\tan x}{1+\cos x}
\label{our_gx}
\end{equation}
and of the conformal factor
\begin{equation}
 \Omega(u,v) = \frac{\cos u \cos v}{4\mu^2}~,
\end{equation}
which together satisfy all necessary conditions if the compactification scale $\mu > 0.38896697...M$ (this bound comes from $h'\ne0$).
The Carter-Penrose diagram with a grid of Schwarzschild coordinates plotted for $\mu=3M$ is shown in Figure \ref{fig_cps}. 
As a consequence of the analytic properties of the transformation we can see in this figure also the regions behind $\scri^\pm$. 
We suppose that in some applications the conformal geometry in a small region behind $\scri^\pm$ may be exploited, e.g. if numerical methods require grid points there. 
The full extension up to the $r\rightarrow 0^-$ available in Figures \ref{fig_cps} and \ref{fig_compacscale} 
does not necessarily have practical applications, but it enables a
direct visual comparison with the well-known conformal embedding of Minkowski spacetime into Einstein static universe. 
In Figure \ref{fig_compacscale}, for larger compactification scales $\mu$ the interior region ($r<2M$) of the black hole on the diagram shrinks, 
but the regions near $i^0$ (both the physical one and those beyond $\scri^\pm$) resemble more and more the compactified Minkowski spacetime 
(see, e.g, the behavior of slices $t=\rm const.$ near $i^0$).

Another common feature -- worldline $r=0^-$ touching the physical region `from behind' $\scri$ at $i^0$ --
illustrates the fact, that (logarithmic) singularity is present at $i^0$ when $M>0$ (see, e.g., \cite{schmidt,herberthson} 
for its detailed description).

\section{Concluding remarks}
\label{SecConcl}
The main visual difference between the compactified diagrams of the Schwarzschild black hole spacetime of Figures \ref{fig_ck} and \ref{fig_cps} is, 
indeed, the angle $\sigma$ at which the singularity $r=0$ approaches $i^\pm$. In both cases the
coordinate transformations have the form of Eq. (\ref{fruv3}) so the angle $\sigma$ is given by the ratio of derivatives 
$h'(0)/h'(\pi/2)$.
The straight shape of the singularity $r=0$ in Figure \ref{fig_ck} is thus related to the very symmetric form of the transformation function 
$h_{\rm MTW}(x)=2M \ln \tan (x)$ (this substitution turns (\ref{fruv3}) into (\ref{g_MTW})) for which $h_{\rm MTW}(\pi/2-x) = - h_{\rm MTW}(x)$.
On the other hand the conditions (\ref{hbeh0}) and (\ref{hbehpi2}) for the analytic extension of coordinates through $\scri^\pm$ and $\cal H^\pm$ are different 
and no similar relation applies.

\begin{figure}[h]
\includegraphics[width=0.2\textwidth,bb=0 315 195 487]{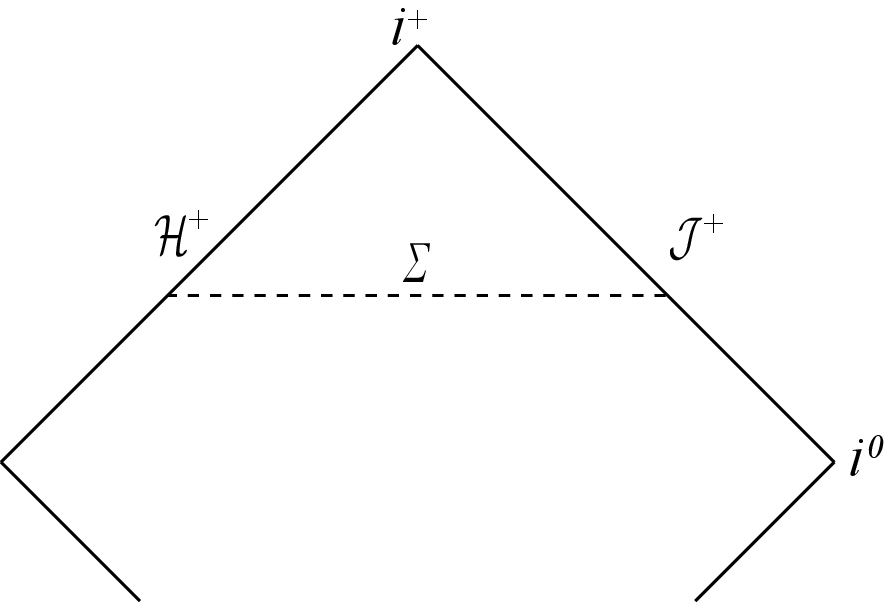}  
\includegraphics[angle=270,width=0.39\textwidth,bb=201 25 554 385]{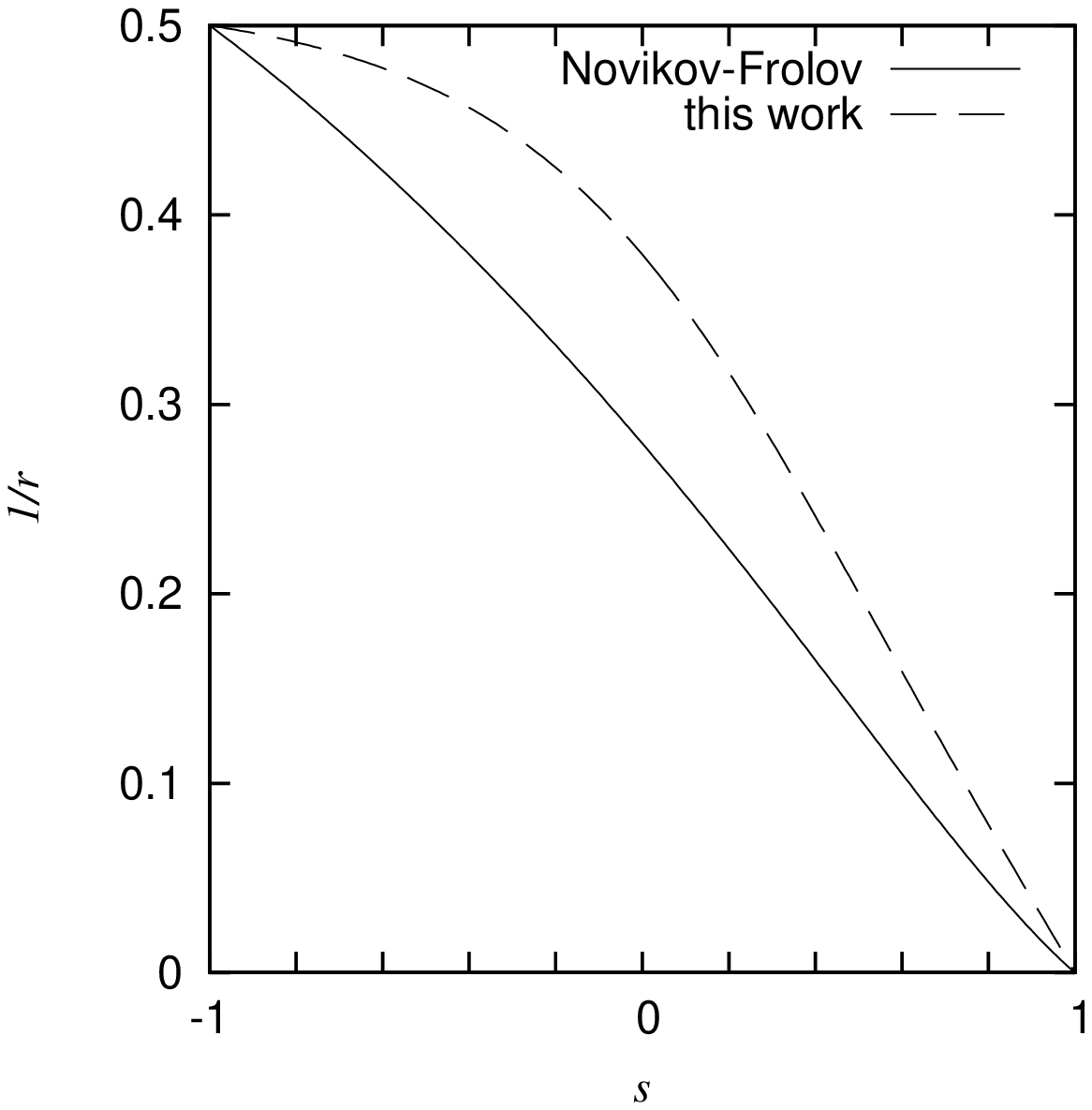}  
\includegraphics[angle=270,width=0.39\textwidth,bb=201 32 554 392]{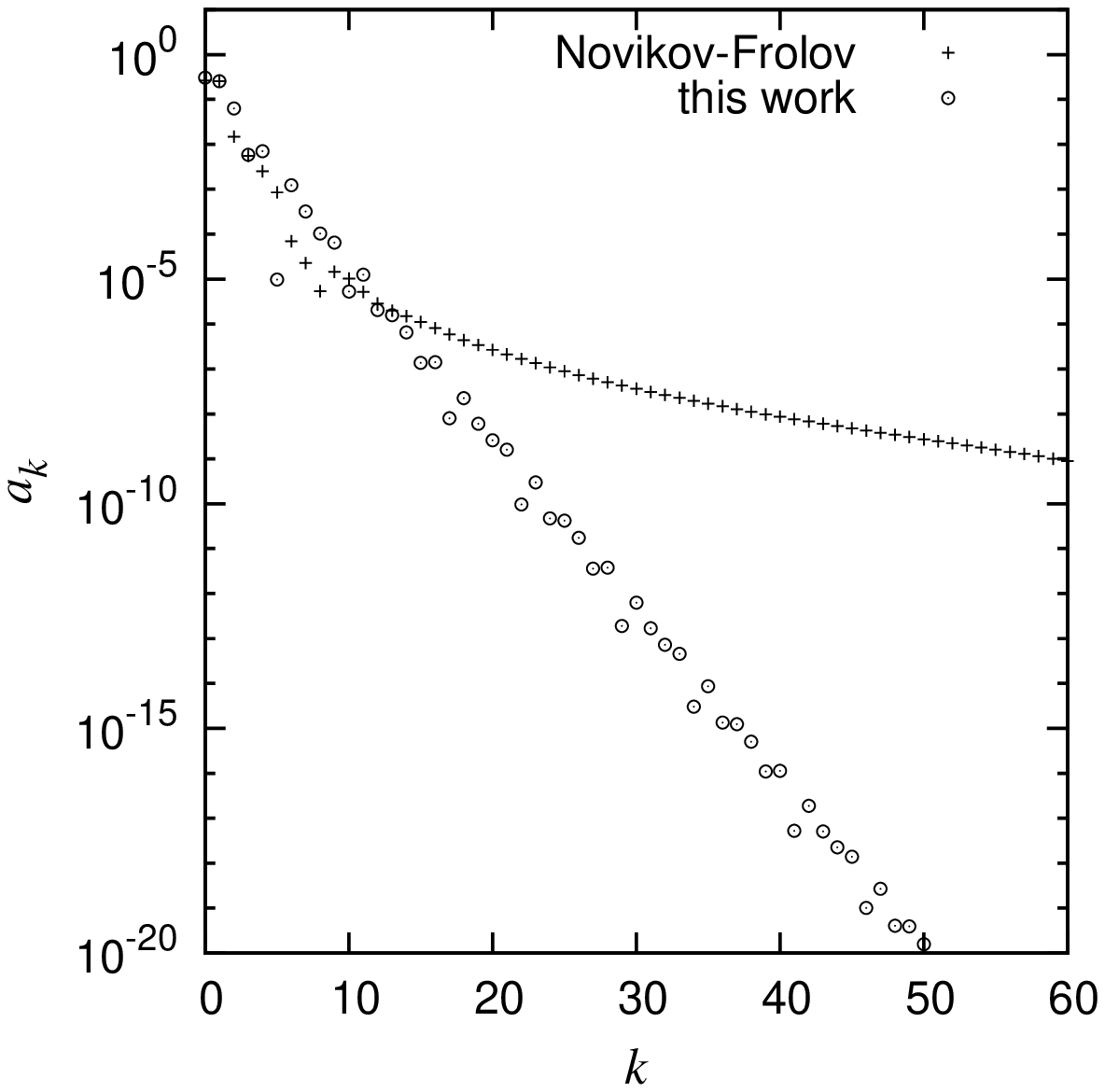}
\\
\begin{center}
\scriptsize
a)~~~~~~~~~~~~~~~~~~~~~~~~~~~~~~~~~~~~~~~~~~~b)~~~~~~~~~~~~~~~~~~~~~~~~~~~~~~~~~~~~~~~~~~~~~~~~~~~c)~~~  
\end{center}
\caption{A comparison of analytic coordinates (\ref{fruv3}) with (\ref{our_gx}) and coordinates suggested in \cite{novikov}.
a) An illustration of the hypersurface connecting horizon and null infinity. 
b) Plot of function $1/r({\mathscr U}(s), {\mathscr V}(s))$ and $1/r(u(s), v(s))$ (see Eqs. (\ref{fruv2}) and (\ref{fruv3}) ) on a linear combination of either coordinates which represents 
radial parameter within the hypersurface. 
c) Coefficients of Chebyshev expansion of functions $1/r({\mathscr U}(s), {\mathscr V}(s))$ and $1/r(u(s), v(s))$.
One can see, that coefficients of $1/r(u(s), v(s))$ decay much faster. See discussion in text
for an example of practical implications.
}
\label{fig_hypsurf}
%
\end{figure}

If we take coordinates (\ref{UVNFdef}) as proposed in \cite{novikov}, the Carter-Penrose diagram would in the physical regions look very similar to the analytic one in Figure \ref{fig_cps}. But as we already mentioned this transformation does not yield an analytic coverage of null infinity. 
Even though the analytic properties of transformations cannot be easily seen from a plot they may still have practical implications. 
As an example, let us consider a situation, where we would like to solve numerically a problem involving an object which spans from horizon $\cal H^+$ to null infinity $\scri^+$, 
e.g., the hyperboloidal hypersurfaces discussed in \cite{husa}. 
Assume also, that the problem can be cast into the form of a differential equation.
The coefficients of this differential equation would contain the Schwarzschild geometry represented by function $1/r$ depending on the compactified coordinates and the regularity of these coefficients would determine the regularity of the solution and, e.g., the behavior of numerical methods used to find this solution. Since various choices of the compactified coordinates yield significant differences in the coordinate dependence of $1/r$, as will be illustrated below, we can  observe differences between the compactified coordinates even for problems formulated completely inside the physical spacetime $r<\infty$.

In Figure \ref{fig_hypsurf}, the analytic coordinates (\ref{fruv3}) with (\ref{our_gx}) and the coordinates suggested in \cite{novikov} are compared.
First we show a plot of function $1/r(s)$ when parameter $s$ linearly advances along a straight line from horizon to null infinity either in compactified coordinates ${\mathscr U}, {\mathscr V}$ or $u, v$ (as illustrated in Figure \ref{fig_hypsurf}a).
Indeed, one cannot distinguish on this plot which of the functions behaves better.
To test this, both functions are decomposed
into the Chebyshev series $1/r(s)=\sum a_k T_k(s)$ and the absolute values of the coefficients $a_k$ are
plotted as a function of $k$ in Figure \ref{fig_hypsurf}c. One can see that for analytic compactification the coefficients decay much faster (exponentially). This indicates some of the
practical implications of using analytic Carter-Penrose compactifications:
numerical methods which require (or take advantage of) good analytic
properties of the function involved would provide better (or faster) results. 
\\
{\it Acknowledgements.}
We are grateful for support from the grants GAUK 606412, GA\v{C}R 205/09/H033, SVV-267301 (J.H.), and GA\v{C}R 202/09/0772 (T.L.).

\appendix
\section{Asymptotically flat spacetimes}
\label{AppA}
The standard way to introduce the notion of asymptotically flat spacetimes refers to the Minkowski spacetime.
There the coordinates $\bar{u}_M,\bar{v}_M$ are usually given by the transformation
\begin{equation}
 2r_M= \tan \bar{v}_M - \tan \bar{u}_M~,~~~2t_M= \tan \bar{v}_M + \tan \bar{u}_M
\label{minkkompakt}
\end{equation}
which  changes the usual Minkowski line element in spherical coordinates $ds_M^2=-dt_M^2+dr_M^2+r_M^2\; d\omega^2 $ with $d\omega^2=d\theta^2+\sin^2\theta d\phi^2$ into
\begin{equation}
 \overline{ds}{}_M^{\,2} = \frac{1}{\cos^2 \bar{u}_M \cos^2 \bar{v}_M} \left( - d\bar{u}_M d\bar{v}_M  + \frac{1}{4}\sin^2(\bar{v}_M-\bar{u}_M) \; d\omega^2 \right).
 \label{dsmink}
\end{equation}
The infinities of Minkowski spacetime $M$ appear on the boundary $\scri:\bar{u}_M=-\pi/2~ \vee~ \bar{v}_M=\pi/2$, and depending upon the character of geodesics which end at those points, 
the spacelike infinity ($i^0$), the future and past null infinity ($\scri^\pm$), and the future and past time-like infinity ($i^\pm$) can be distinguished. 
Then, using the conformal factor $\Omega_M = \cos \bar{u}_M \cos \bar{v}_M$ which can be clearly identified in (\ref{dsmink}), 
we obtain a larger manifold $\widetilde M$ with metric ${\widetilde{ds}}{}_M^2 = \Omega^2_M ~\overline{ds}{}_M^{\,2}$
regular on $\scri$.
The coordinates $\widetilde{u}_M=\bar{u}_M, \widetilde{v}_M=\bar{v}_M$ are no longer restricted to $[-\pi/2,\pi/2]\times[-\pi/2,\pi/2]$. 

The choice of the compactified coordinates and the conformal factor above are restricted by 
the requirements present in the definition of the class of the {\it asymptotically flat spacetimes} \cite{wald,ashtekar,penrose}:
the coordinates and the conformal factor $\Omega$ must lead to the conformally related metric
\begin{equation}
\widetilde{ds}{}^2 = \Omega^2 ~{ds}\,{}^2
\label{dsOmegads}
\end{equation}
regular at null infinity and the conformal factor must vanish at infinity with the leading terms in its power expansion near $\scri$ being 
\begin{eqnarray}
 \Omega({\scri^\pm}) &= 0~,~\widetilde \nabla_\mu \Omega({\scri^\pm}) \ne 0~,&~\label{dOmega}\\
 \Omega({i^0}) &= 0~,~\widetilde \nabla_\mu \Omega({i^0}) = 0~,&~\widetilde \nabla_\mu \widetilde \nabla_\nu \Omega({i^0}) = 2 \widetilde g_{\mu \nu}({i^0}).\label{ddOmega}
\end{eqnarray}
Note that a simple choice of the conformal factor $\Omega = 1/r$ does not satisfy (\ref{ddOmega}) despite the fact that such a choice is sometimes suggested.

\section*{References}

\bibliography{HL}

\end{document}